\documentclass[a4paper,showpacs,nofootinbib]{revtex4-1}
\usepackage[utf8]{inputenc}
\usepackage{graphicx,bm,color}
\usepackage{slashed,verbatim}
\usepackage{subfigure}
\usepackage[colorlinks=true,linktocpage=true,linkcolor=blue,citecolor=blue]{hyperref}

\date{\today}

\begin{document}

\title{Vector meson spectral function and dilepton rate in the presence of strong entanglement
effect between the chiral and the Polyakov loop dynamics}
\author{Chowdhury Aminul Islam, Sarbani Majumder, Munshi G. Mustafa}
\affiliation{Theory Division, Saha Institute of Nuclear Physics, 1/AF Bidhan Nagar, Kolkata-700064, India }

\begin{abstract}
In this  work we have re-explored  our earlier study on the vector meson spectral function and  
its spectral property in the form of dilepton rate in a two-flavour Polyakov loop extended Nambu$\textendash$Jona-Lasinio 
(PNJL) model in presence  of a strong entanglement between the chiral and Polyakov loop dynamics. 
The entanglement considered here is generated through the four-quark scalar type interaction in which the coupling 
strength depends on the Polyakov loop and runs with temperature and chemical potential. The entanglement effect 
is also considered for the four-quark vector type interaction in the same manner. We observe that the entanglement 
effect relatively enhances the color degrees of freedom  due to the running
of the both scalar and vector couplings. This modifies the vector meson spectral function  and thus  
the spectral property such as the dilepton production rate in low invariant mass also gets modified.
\end{abstract}


\maketitle
\section{Introduction}

Quantum chromodynamics (QCD) is a theory of strong interaction
that accounts for the rich phenomenology of hadronic and
nuclear physics.  However, the theory is not yet very well understood  because of its
nonperturbative nature.  Ongoing ultra-relativistic heavy-ion collision experiments at RHIC BNL and LHC CERN
have indicated the formation of  quark-gluon plasma (QGP), a deconfined state of matter, transformed from hadronic
matter at very high temperatures and/or densities as predicted by asymptotic freedom of QCD.
The QCD phase diagram is not only essential for understanding the phenomena in the laboratory experiments 
involving relativistic heavy-ion collisions but also for the natural phenomena such as compact stars and 
the early universe. 

The phase diagram of hot and/or dense system of quarks and gluons predicted by the QCD has invited a 
lot of serious theoretical  investigations for last few decades. The first prototype of the QCD phase 
diagram was conjectured in \cite{Cabibbo:1975ig} where it looked very simple; with the passage of time
more and more investigations culminated in a very complicated looking phase diagram with many exotic 
phases \cite{Fukushima:2010bq}. Nevertheless, the interest mainly revolved around  two phase 
transitions - one is the chiral phase transition  and the other one is the deconfinement transition.
If they do not coincide, exotic phases such as the constituent quark phase~\cite{Cleymans:1986cq,Kouno:1988bi} or the
quarkyonic phase~\cite{McLerran:2007qj,Hidaka:2008yy} may occur. So, an important question on the QCD thermodynamics 
is whether the chiral symmetry restoration and the confinement-to-deconfinement transition happen simultaneously or not. 
We note that chiral and deconfinement transitions are conceptually two distinct phenomena. Though lattice QCD simulation 
has confirmed that these two transitions occur at the same temperature \cite{Fukugita:1986rr} or almost at the 
same temperature \cite{Aoki:2006br}. Whether this is a mere coincidence or some dynamics between the two phenomena 
are influencing each other is not understood yet and is matter of intense current research exploration. 

To understand the reason behind this  coincidence a conjecture has been proposed in 
the article \cite{Sakai:2010rp}  through a strong correlation or entanglement between the 
chiral condensate ($\sigma$) and the Polyakov loop expectation value ($\Phi$) 
within the Polyakov loop extended Nambu$\textendash$Jona-Lasinio (PNJL) model. Usually, in 
PNJL model, there is 
a weak correlation between the chiral dynamics $\sigma$ and the confinement-deconfinement 
dynamics  $\Phi$ that is in-built through the covariant derivative between quark and gauge fields.
With this kind of weak correlation the coincidence between the chiral and deconfinement 
crossover transitions~\cite{Sakai:2009dv,Ghosh:2006qh,Mukherjee:2006hq,Ghosh:2007wy,Ratti:2005jh,Deb:2009ng}
can be described but it requires  some fine-tuning of parameters, inclusion of the scalar type
eight-quark interaction for zero chemical potential $\mu$ and the vector-type four-quark interaction
for imaginary $\mu$. This reveals that there may be a stronger correlation 
between $\Phi$ and $\sigma$ than that in the usual PNJL model associated through the
covariant derivative between quark and gauge fields. Also, some recent analyses~\cite{Braun:2009gm,Kondo:2010ts} 
of the exact renormalization-group (ERG) equation~\cite{Wetterich:1992yh} suggest a strong entanglement 
interaction between $\Phi$ and $\sigma$ in  addition to the original entanglement through the covariant
derivative. Based on this the two-flavor PNJL model is further generalized~\cite{Sakai:2010rp} 
by considering the effective four-quark scalar type interaction with the coupling strength that 
depends on the Polyakov Loop (PL) field $\Phi$. The effective vertex in turn generates entanglement interaction 
between $\Phi$ and $\sigma$. Such generalization  of the PNJL model is known as
Entangled-PNJL (EPNJL) model~\cite{Sakai:2010rp}. This EPNJL model has been used to study the location of 
the tricritical point at real isospin chemical potential~~\cite{Sakai:2010rp} and
on the location of the critical endpoint at real quark-number chemical potential~\cite{Sakai:2010rp,Friesen:2014mha,Sugano:2014pxa}.
It has also been used to study~\cite{Restrepo:2014fna} the effect of dynamical generation of a repulsive vector contribution to the quark pressure.
The EPNJL model has further been generalized to the three-flavor phase diagram~\cite{sasaki:2011} as a function of light- and 
strange-quark masses for both zero and imaginary quark chemical potential.

It is well known that many of the hadron properties are encoded in the correlation function and its spectral
representation. The properties of the vector current correlation function and its spectral representation 
in the deconfined phase have been studied to understand the nonperturbative effect on the vector 
current spectral properties, e.g., the dilepton production rate in lattice QCD (LQCD) 
framework~\cite{Ding:2011}.
Recently, in  PNJL model that takes into account nonperturbative effects like chiral and confinement dynamics, 
we have analysed~\cite{Islam:2014sea} the effect of isoscalar-vector 
interaction on the 
vector meson spectral function and its various spectral properties ({\it viz.}, dilepton production rate\footnote{We also note that 
both the dilepton and real photon rate have been computed in a matrix model of QGP by considering only the confinement 
effect~\cite{Gale:2014dfa,Hidaka:2015ima} and taking into account both the confinement and chiral symmetry breaking 
effects~\cite{Satow:2015oha}.} and the quark number susceptibility (QNS) 
associated with the conserved density fluctuation) in a hot and dense medium.
In present  article, we consider the idea of the EPNJL model in which the effective vertex  generates a strong entanglement interaction 
between the chiral condensate $\sigma$ and the Polyakov loop $\Phi$ to re-explore  the vector spectral function and the spectral 
property such as the dilepton production rate previously  studied in \cite{Islam:2014sea}. Because of this strong entanglement 
between $\Phi$ and $\sigma$, the coupling strengths run with the temperature and chemical potential. First we study the characteristics 
of mean fields with various constraints: with and without the isoscalar-vector interaction in both PNJL and EPNJL models. Then we further 
demonstrate the effect of the entanglement on vector meson spectral function and dilepton rate.

The paper is organized as follows: in sec.~\ref{Eff_QCD_model} we briefly outline the usual PNJL model and extend it with the 
entanglement effect, namely the EPNJL model. In sec.~\ref{spect} we write the expression for the vector spectral function
and its various spectral properties following our earlier calculation in Ref.~\cite{Islam:2014sea}. In sec~\ref{results} we
discuss our results and finally we conclude in sec.~\ref{concl}.
\section{EFFECTIVE QCD MODEL}
\label{Eff_QCD_model}

\subsection{PNJL Model}
\label{pnjl}

We start with the two flavour PNJL model Lagrangian with isoscalar-vector interaction~\cite{Islam:2014sea}

\begin{eqnarray}
 {\mathcal L}_{\rm PNJL} &=& \bar{\psi}(i\slashed D-m_0+\gamma_0\mu)\psi +
\frac{G_S}{2}[(\bar{\psi}\psi)^2+(\bar{\psi}i\gamma_5\vec{\tau}\psi)^2]
- \frac{G_V}{2}(\bar{\psi}\gamma_{\mu}\psi)^2\nonumber\\
&-& {\mathcal U}(\Phi[A],\bar{\Phi}[A],T),
\label{pnjl_lagrangian}
\end{eqnarray}
where $\psi$ denotes the two flavour quark field,  $m_0 = $diag$(m_{u},m_{d})$ with $m_{u}=m_{d}$ and $\vec{\tau}$'s
are Pauli matrices. $D^\mu=\partial^\mu-ig{\mathcal A}^\mu_a\lambda_a/2$, ${\mathcal A}^\mu_a$ being the $SU(3)$ 
background fields, $\lambda_a$'s are the Gell-Mann matrices and $g$ is the gauge coupling; $ G_{S}$ and $G_{V}$ 
are, respectively, the coupling constants of local scalar type four-quark interaction and isoscalar-vector 
interaction, which do not run. $\mathcal U$ is the Polyakov potential that depends on the Polyakov Loop $\Phi$ 
and it's charge conjugate $\bar\Phi$.

The corresponding thermodynamic potential is obtained as 
\begin{eqnarray}
 {\Omega}_{\rm{PNJL}} &=& {\mathcal U}(\Phi,{\bar \Phi},T) + \frac{ G_S}{2} \sigma^2 -\frac{ G_V}{2} n^2 
 \nonumber\\
 &-&2N_fT\int \frac{d^3p}{(2\pi)^3} \ln \left[1+ 3\left(\Phi +{\bar \Phi}
 e^{-(E_p-\tilde \mu)/T} \right)e^{-(E_p-\tilde \mu)/T} + e^{-3(E_p-\tilde \mu)/T} \right ]  \nonumber \\
 &-&  2N_fT\int \frac{d^3p}{(2\pi)^3} \ln \left[1+ 3\left({\bar \Phi} + \Phi
 e^{-(E_p+\tilde \mu)/T} \right)e^{-(E_p+\tilde \mu)/T} + e^{-3(E_p+\tilde \mu)/T} \right ] \nonumber \\ 
 &-&\kappa T^4 \ln[J(\Phi,{\bar \Phi})] 
  -2N_fN_c\int_{\Lambda}\frac{d^3p}{(2\pi)^3}E_p\ .
 \label{eq.thermo_pot_pnjl}
\end{eqnarray}
Here $E_p=\sqrt{{\vec p}^2+M_f^2}$ is the energy of a quark with flavor $f$ having constituent mass or 
the dynamical mass $M_f$ and $\Lambda$ is a finite three momentum cut-off. This $M_f$ and the effective 
quark chemical potential $\tilde\mu$ are related with the scalar ($\sigma$) and vector ($n$) condensates as

\begin{equation}
 M_f=m_0-G_S\sigma , \label{eq.massgap}
\end{equation}
and
\begin{equation}
 \tilde{\mu}=\mu-G_V n, \label{eq_mu_tilde}
\end{equation}
respectively. It is noteworthy that the value of $G_S$ along with the current quark mass ($m_0$) and the three momentum cutoff ($\Lambda$) 
are fixed in the NJL model itself to reproduce some zero temperature results namely pion mass, pion decay constant and the quark 
condensate~\cite{Ratti:2005jh}. But the value~\footnote{Some efforts have also been made to estimate the value of 
$G_V$ mainly by fitting lattice data through two-phase model, which is not very conclusive. Interested readers are referred to  
Refs.~\cite{Steinheimer:2014kka,Sugano:2014pxa} and references therein.} of $G_V$ cannot be fixed within the 
formalism of this model, since its value is to be fixed
using the mass of the $\rho$ meson which is beyond the maximum energy scale $\Lambda$ of the model. So we keep $G_V$ as free parameter 
and consider different choices as $G_V = x\times G_S$, where $x$ is a multiplicative factor chosen in the range from $0$ to $1$.  
We use $\mathcal U$ from reference \cite{Ratti:2005jh} which is fitted to lattice QCD in pure gauge theory 
at finite temperature and is given by

\begin{equation}
 \frac{{\mathcal U}(\Phi,\bar{\Phi},T)}{T^4} = 
    -\frac{b_2(T)}{2}\Phi\bar{\Phi} -
    \frac{b_3}{6}(\Phi^3+{\bar{\Phi}}^3) +
    \frac{b_4}{4}(\bar{\Phi}\Phi)^2,
\label{eq.potential}
\end{equation}
with
\begin{equation}
 b_2(T) = a_0 + a_1\left(\frac{T_0}{T}\right) + a_2\left(\frac{T_0}{T}\right)^2 +
    a_3\left(\frac{T_0}{T}\right)^3. \nonumber\\
\end{equation}
Values of different coefficients $a_0,\ a_1,\ a_2,\ a_3,\ b_3$ , $b_4$ and $\kappa$ have been 
tabulated in~\cite{Islam:2014sea}. The Vandermonde term $J(\Phi,{\bar \Phi})$ is given as \cite{Ghosh:2007wy}
\begin{equation}
J[\Phi, {\bar \Phi}]=\frac{27}{24\pi^2}\left[1-6\Phi {\bar \Phi}+
4(\Phi^3+{\bar \Phi}^3)-3{(\Phi {\bar \Phi})}^2\right].
\end{equation}
In the pure gauge theory the Polyakov potential is fitted to lattice QCD that yields a first order phase transition at $T_0=270$ MeV. 
With this value of $T_0$ for zero chemical potential we get, for 2-flavour case, almost a coincidence between the chiral and deconfinement 
transitions\footnote{We note that the chiral transition temperature $T_\sigma$ is obtained from the peak position of 
the  $\partial\sigma/\partial T$ whereas the deconfinement transition temperature 
$T_\Phi$ is that from the  $\partial \Phi/\partial T$.} 
($T_\sigma = 233$ MeV and $T_\Phi=228$ MeV). So the two transitions almost coincide [e.g., Fig.~\ref{Tsigma_Tphi_PNJL}]
but at a value higher than the range provided by 
the 2-flavour\footnote{It is worth mentioning here that the chiral transition temperature is found 
to be $ T_c= (154\pm 9)$ MeV in the recent (2+1) flavour LQCD computations by HotQCD collaboration~\cite{Bazavov:2011nk}.
In (2+1) flavor QCD the chiral order parameter contains both 
the light quark condensate and the strange quark condensate. 
Only the former is used to define the chiral transition temperature, as the strange condensate varies
very smoothly~\cite{Bazavov:2013yv}. Now, the behaviour of the light quark condensate
in (2+1) flavour and 2-flavour QCD will be similar
if the light quark masses are similar but  will be
different at quantitative level as it leads to two different chiral transition 
temperatures simply because one has {\it two different scales} in the theory.   
The value $ T_c= (154\pm 9)$ MeV was extracted~\cite{Bazavov:2011nk}  entirely in reference to the chiral 
phase transition for (2+1) flavour QCD. Further, we also note that the Wuppertal-Budapest 
collaboration~\cite{Borsanyi:2010bp} has also extracted three 
somewhat different values of $T_c$ ranging from $147$ MeV
to $157$ MeV, depending on the chiral observables considered for the purpose.
Since we  restrict our calculation only to 2-flavour case, we stick to the corresponding  
$T_c=(173\pm8$) MeV as extracted for 2-flavour case in LQCD simulation\cite{Karsch:2001cy,Karsch:2000kv}.} 
lattice QCD \cite{Karsch:2001cy,Karsch:2000kv} which is  
$T_\sigma\approx T_\Phi\approx (173\pm8)$ MeV. In 
Ref~\cite{Ratti:2005jh} the value of $T_0$ was changed to 190 MeV but keeping all the other parameters  same and obtained 
a lower value of $T_\sigma$ ($\approx200$ MeV) and $T_\Phi$ ($\approx170$ MeV) [e.g., Fig.~\ref{Tsigma_Tphi_PNJL(T0=190)}]. 
Taking the average of the two while defining $T_c$ 
gives a value almost within the range provided by the lattice QCD but then the coincidence is lost.
In this article we work with the same Polyakov potential but the entanglement between the chiral and deconfinement mechanism 
is introduced in the next subsec.~\ref{epnjl}.

\subsection{EPNJL}
\label{epnjl}

The PNJL model has a weak correlation between the chiral ($\sigma$) and the deconfinement  ($\Phi$ and $\bar \Phi$) dynamics through
the covariant derivative between the quark and  the gauge fields.  In addition to this  there may be
a strong entanglement interaction between $\Phi$ and $\sigma$  as suggested by  some recent analyses~\cite{Braun:2009gm,Kondo:2010ts} 
of the ERG equation~\cite{Wetterich:1992yh}. Based on this the two-flavor PNJL model is 
further generalized by considering the effective scalar~\cite{Sakai:2010rp} and vector~\cite{Sugano:2014pxa}
type four-quark interaction with the coupling strengths that 
depend on the Polyakov field $\Phi$. The Lagrangian in EPNJL will be the same as that in (\ref{pnjl_lagrangian}) except that now 
the coupling constants $G_S$ and $G_V$ will be replaced by the effective ones $\tilde G_S(\Phi)$ and $\tilde G_V(\Phi)$. The effective 
vertices $\tilde{G}_S(\Phi)$ and  $\tilde{G}_V(\Phi)$ in turn generates entanglement interaction 
between $\Phi$ and $\sigma$ and their forms are chosen~\cite{Sakai:2010rp,Sugano:2014pxa}
to preserve chiral and $Z_3$ symmetry as given by
\begin{equation}
 \tilde{G}_S(\Phi)= G_S[1-\alpha_1\Phi\bar\Phi-\alpha_2(\Phi^3+\bar\Phi^3)]  , \label{entangle_Gs}
\end{equation}
and
\begin{equation}
 \tilde{G}_V(\Phi)= G_V[1-\alpha_1\Phi\bar\Phi-\alpha_2(\Phi^3+\bar\Phi^3)]  . \label{entangle_Gv}
\end{equation}
We  note that for $\alpha_1 =\alpha_2=0$, 
$\tilde{G}_S(\Phi)= G_S$ and $\tilde{G}_V(\Phi)= G_V$, the EPNJL model reduces to PNJL model. Also
at $T=0$, $\Phi={\bar \Phi}=0$ (confined phase), then $\tilde{G}_S= G_S$ and $\tilde{G}_V= G_V$. Due to the reason already mentioned in 
the previous subsection, here again the strength of the vector interaction is taken in terms of the value of $G_S$ as $G_V=x\times G_S$, 
which on using (\ref{entangle_Gv}) reduces to 
\begin{equation}
\tilde{G}_V(\Phi)=x \times G_S[1-\alpha_1\Phi\bar\Phi-\alpha_2(\Phi^3+\bar\Phi^3)]=x \times \tilde{G}_S(\Phi) . \label{entangle_Gv_tilde}
\end{equation}
Now 
in EPNJL model, $\alpha_1$ and $\alpha_2$ are two new parameters, which are
to be fixed from the lattice QCD data. The thermodynamic potential $\Omega_{\rm{EPNJL}}$ in EPNJL model can be obtained 
from (\ref{eq.thermo_pot_pnjl}) by replacing
$G_S$ with $\tilde{G}_S(\Phi)$ and $G_V$ with $\tilde{G}_V(\Phi)$.
For the EPNJL model we take same values of the parameters 
as those in PNJL model~\cite{Islam:2014sea} except the value of $T_0$, which is taken as 190 MeV. 
Then we fix the values of parameters $\alpha_1$ and $\alpha_2$  so as  to reproduce the coincidence of chiral and deconfinement 
transitions within the range given by lattice QCD data at zero chemical potential \cite{Karsch:2001cy,Karsch:2000kv} and it 
is found that ($\alpha_1$,~$\alpha_2$)= (0.1, 0.1). We further  mention that the coincidence of $T_\sigma$ and $T_\Phi$ are preserved 
[e.g., Fig.~\ref{Tsigma_Tphi_epnjl}]
within the parameter region $\alpha_1$, $\alpha_2$ $\approx$ 0.10 $\pm$ 0.05. Note that the values 
$\alpha_1$ and $\alpha_2$ in our model differ from that of Ref.~\cite{Sakai:2010rp} because of the choice of
different Polyakov loop potential. We chose the form of the potential as given in Refs.~\cite{Ratti:2005jh} 
whereas that used in Ref.~\cite{Sakai:2010rp}  is taken from Ref.~\cite{Roessner:2006xn}. It is also noteworthy that the two forms 
of Polyakov Loop potentials are consistent with each other in the validity domain of the model~\cite{Fukushima:2008wg}.

\section{Vector meson spectral function and spectral properties}
\label{spect}

\subsection{Resummed vector meson spectral function in ring summation}

The resummed vector meson spectral function in presence of isoscalar-vector interaction within ring
approximation is in Ref.~\cite{Islam:2014sea}
\begin{equation}
\sigma_V(\omega, {\vec q})=\frac{1}{\pi}\Big[{\rm Im}C_{00}(\omega, {\vec q})-{\rm Im}C_{ii}(\omega, {\vec q})\Big ],
\label{eq.spectral_resum}
\end{equation}
where $Q\equiv(\omega, {\vec q})$, the four momentum of the vector meson. The imaginary part of the temporal component 
of the resummed correlator $C_{00}$ is given 
as
\begin{equation}
 {\rm Im C_{00}}=\frac{{\rm Im \Pi_{00}}}
 {\Big[ 1- {\tilde G}_V(\Phi) \Big(1-\frac{\omega^2}{{q}^2}\Big){\rm Re \Pi_{00}}\Big]^2+
 \Big[{\tilde G}_V(\Phi) (1-\frac{\omega^2}{{q}^2})
 {\rm Im \Pi_{00}}\Big ]^2},\label{eq.C00}
\end{equation}
and imaginary part of the spatial vector correlator is given as
\begin{eqnarray}
 {\mbox{Im}}C_{ii}=\frac{{\rm{Im}}\Pi_{ii}-\frac{\omega^2}
 {{q}^2}{\rm{Im}}\Pi_{00}}
 {\left[1+\frac{{\tilde G}_V(\Phi)}{2}{\rm{Re}}\Pi_{ii}-\frac{{\tilde G}_V(\Phi)}{2}
 \frac{\omega^2}
 {{q}^2}{\rm{Re}}\Pi_{00}\right]^2
 +\frac{{\tilde G}_V^2(\Phi)}{4}\Big[{{\rm{Im}}\Pi_{ii}-\frac{\omega^2}
 {{q}^2}\rm{{Im}}\Pi_{00}}\Big]^2}
 +  \frac{\omega^2}{{q}^2}{\rm Im}C_{00}. \label{eq.CTprime}
\end{eqnarray}
The various expressions for one-loop self-energies, $\Pi_{00}$ and $\Pi_{ii}$, are explicitly
computed in our earlier work in~\cite{Islam:2014sea}.

\subsection{Vector Spectral Function and Dilepton Rate}

 The  vector meson spectral function, $\sigma_V$, and the differential
dilepton production rate are related  as
\begin{equation}
  \frac{dR}{d^4xd^4Q}  =\frac{5\alpha^2} {54\pi^2} \frac{1}{M^2} 
\ \frac{1}{e^{\omega/T}-1} \ \sigma_V(\omega, {{\vec q}}) \ , \label{eq.rel_dilep_spec}
\end{equation}
where the invariant mass of the lepton pair is $M^2=\omega^2-q^2$ and 
$\alpha$ is the fine structure constant.

\section{Results}
\label{results}

\subsection{Mean Fields}
\subsubsection{Without the isoscalar-vector interaction ($G_V=0$)}

\begin{figure}[hbt]
\subfigure[]
{\includegraphics[scale=0.8]{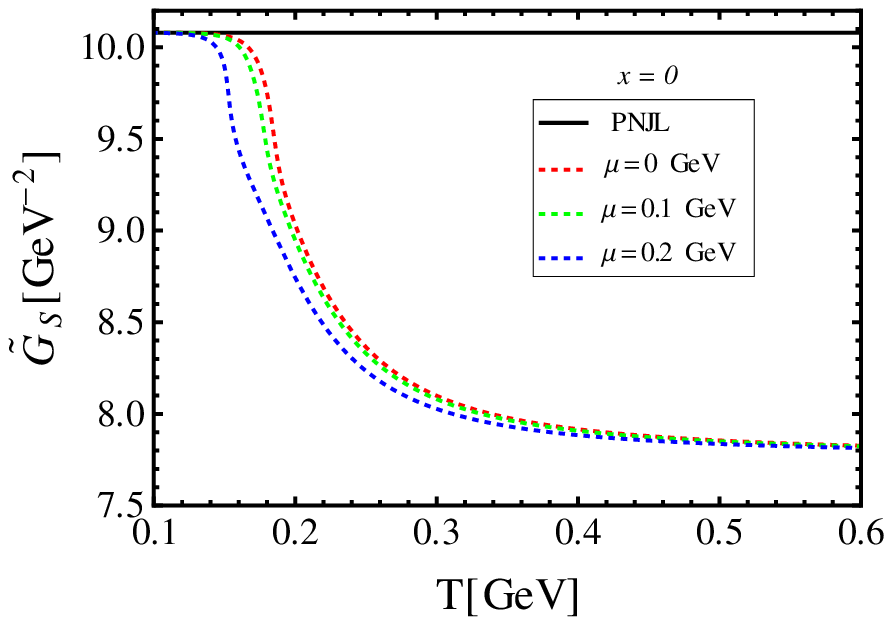}
\label{couplings_Gv0}}
\subfigure[]
{\includegraphics[scale=0.785]{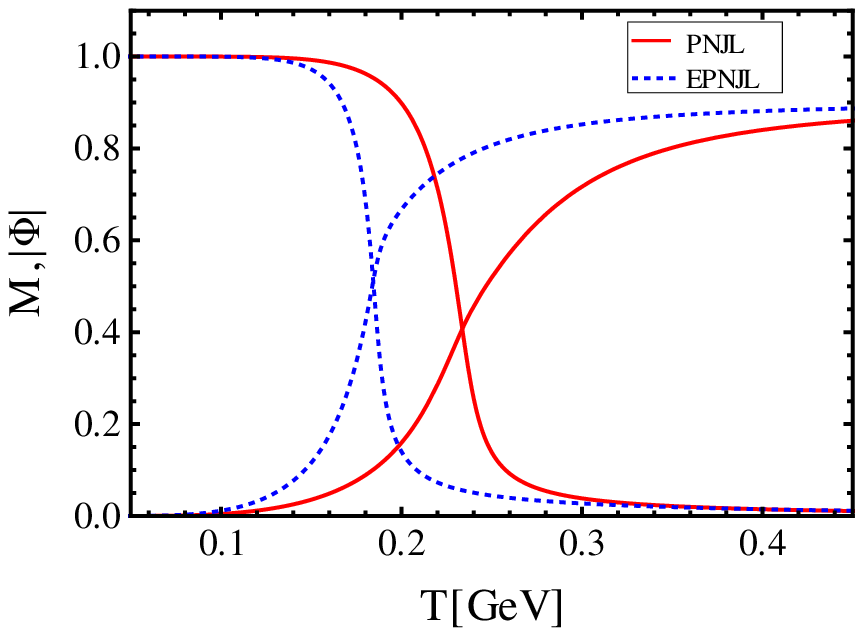}
\label{mass_fields_Gv0_Mu0}}
\caption{ Variation of (a) scalar type four-quark coupling strength ${\tilde G}_S(\Phi)$ with temperature $T$ 
for different values of $\mu$ and (b) the constituent quark mass scaled with its zero temperature value and Polyakov Loop fields with 
$T$  for $\mu=0$ for both PNJL (solid lines) and EPNJL (dotted lines) model.}
\label{coupling_mass_fields_Gv0_Mu0}
\end{figure}

The gap equation for the thermodynamic potential is 

\begin{eqnarray}
\frac{\partial \Omega_{(E)PNJL}}{\partial X}=0, \label{gap_equation}.
\end{eqnarray}
The thermodynamic potential is minimized with respect to mean fields $X$; with $X$ 
representing $\sigma,~ \Phi, ~\bar{\Phi}$ and $n$. In this section we compare the variations of the mean fields in PNJL model with that 
of EPNJL one without the effect of isoscalar-vector interaction i.e., $G_V = 0$. As discussed in subsec.~\ref{pnjl}  the scalar 
type four-quark coupling strength ($G_S$)
in NJL/PNJL model is fixed along with three momentum cutoff $\Lambda$ and bare quark mass $m_0$ to reproduce known zero temperature 
chiral physics in the hadronic sector. We note that in principle it should depend on the parameters $T$ and $\mu$ but it is not usually considered in NJL model~\cite{Klevansky:1992qe,Hatsuda:1994pi}. 
However, in PNJL model the PL field ($\Phi$) is related to the temporal gluon 
which should make $G_S$ to depend on $\Phi$. But this dependence is also neglected 
in the same spirit~\cite{Fukushima:2003fw}. So, the value of $G_S$ remains 
fixed as represented by solid line in Fig.~\ref{couplings_Gv0}.

Now we pay attention to the features of EPNJL model in Fig.~\ref{couplings_Gv0}. As soon as one introduces
the $\Phi$ dependence in the scalar coupling strength through~(\ref{entangle_Gs}) in EPNJL model, it ($\tilde{G}_S$) becomes dependent on both 
$T$ and $\mu$. This running is due to  the gap equation in (\ref{gap_equation}), which is solved 
in a self-consistent manner for different mean fields.  As can be seen the increase in $T$ causes  $\tilde{G}_S$ to decrease 
for a given $\mu$ and the decrease becomes faster as one increases $\mu$.
This can be understood from (\ref{entangle_Gs}) as  for a given $T$ if  one increases $\mu$, the 
Polyakov loop fields ($\Phi$ and $\bar \Phi$) increase  and thus  $\tilde{G}_S$ decreases.
Fig.~\ref{mass_fields_Gv0_Mu0} displays the temperature dependence of 
the scaled constituent quark mass and PL fields for both PNJL and EPNJL models at $\mu =0$. Here we  
mention that for $\mu=0$, $\Phi=\bar\Phi=|\Phi|$ \cite{Ratti:2005jh}. It clearly shows 
a considerable change in the chiral condensate $(\sigma=\langle\bar\psi\psi\rangle)$ 
and the Polyakov loop fields in EPNJL model as compared to those in  PNJL model. For nonzero chemical potential 
similar behaviour of $\sigma$ and $\Phi, \bar\Phi$ is also observed. This is obviously 
due to the running of the coupling ${\tilde G}_S$ which is arising due to the entanglement effect as shown in Fig~\ref{couplings_Gv0}.

\begin{figure}[hbt]
\subfigure[]
{\includegraphics[scale=0.63]{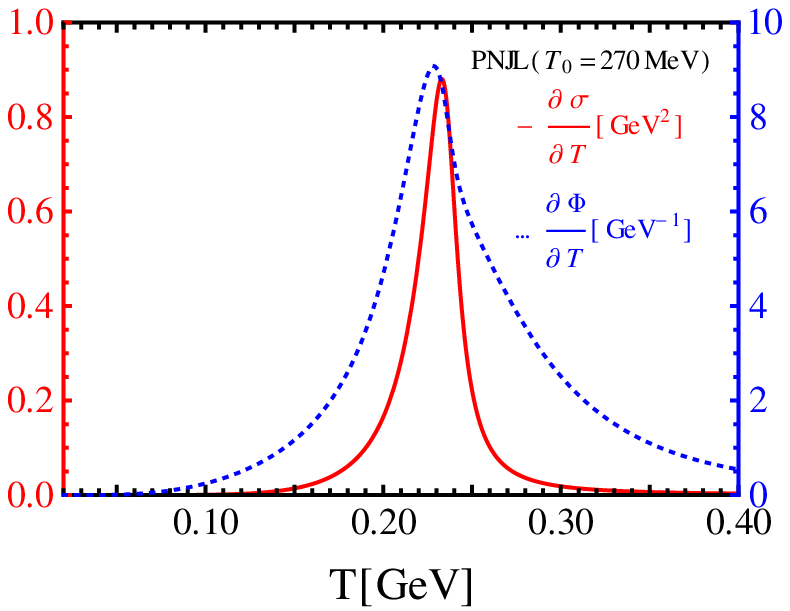}
\label{Tsigma_Tphi_PNJL}}
\subfigure[]
{\includegraphics[scale=0.63]{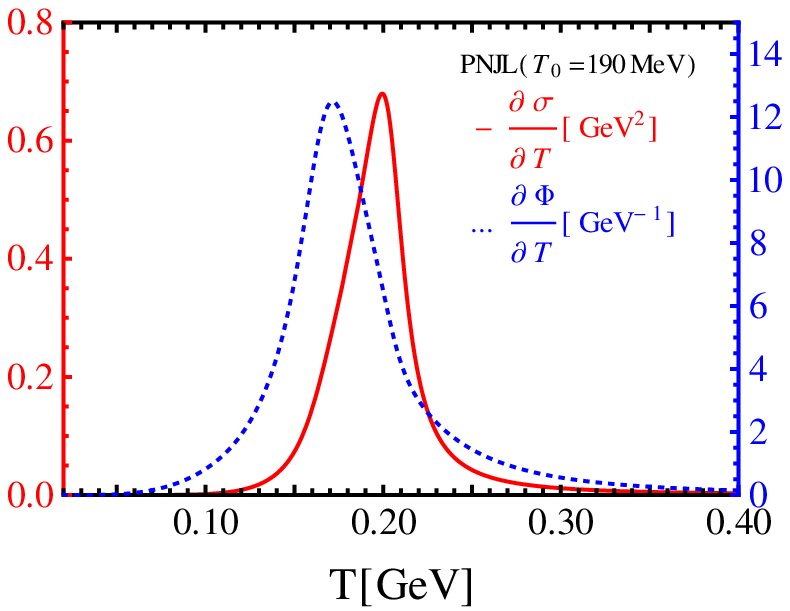}
\label{Tsigma_Tphi_PNJL(T0=190)}}
\subfigure[]
{\includegraphics[scale=0.63]{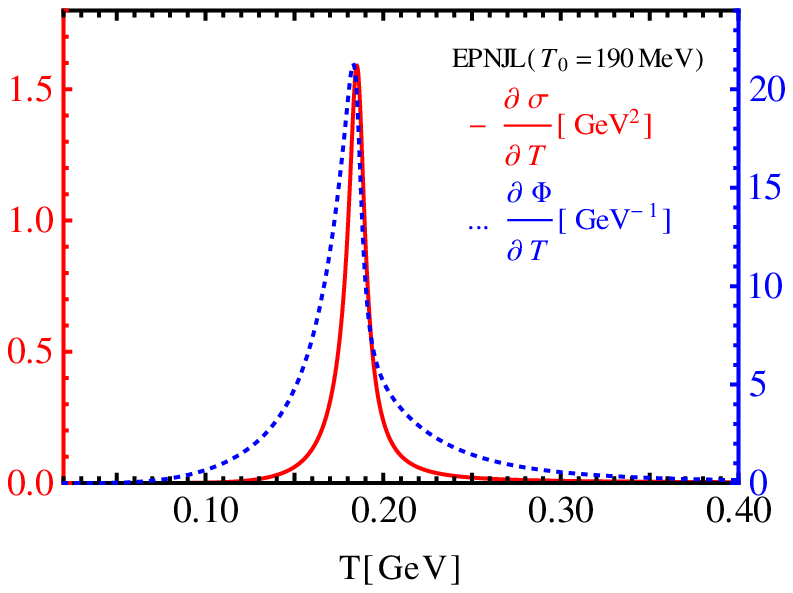}
\label{Tsigma_Tphi_epnjl}}
\caption{Plot of $\partial \sigma/\partial T$ and  
$\partial\Phi/\partial T$  as function of $T$ with $\mu=0$ 
for (a) PNJL model with $T_0=270$ MeV in Ref.~\cite{Ghosh:2007wy,Ratti:2005jh},
(b) PNJL model with $T_0=190$ MeV in Ref.~\cite{Ratti:2005jh} and
(c) present calculation in EPNJL model with $T_0=190$ MeV. For details it is referred to text 
in subsecs.~\ref{pnjl} and ~\ref{epnjl}, respectively.}
\label{Tsigma_Tphi_Gv0_Mu0}
\end{figure}

Fig.~\ref{Tsigma_Tphi_Gv0_Mu0} displays the variations of $\frac{\partial \sigma}{\partial T}$ and 
$\frac{\partial\Phi}{\partial T}$ with the temperature at $\mu=0$ for various model conditions as discussed 
in subsecs.~\ref{pnjl} and ~\ref{epnjl} in details. We note that  $T_\sigma$ and $T_\Phi$ coincide for EPNJL model 
at $\approx 184$ MeV (e.g, Fig.~\ref{Tsigma_Tphi_epnjl}), which is almost within the range, $T_c=(173\pm8)$ MeV,  
given by the two flavour lattice  QCD~\cite{Karsch:2001cy,Karsch:2000kv}.

We note that once $\mu$ is introduced in the system the transition temperatures 
(both chiral and deconfinement) get reduced, which is expected. 
Now for a  given $T$ and $\mu\ne 0$,  $\Phi \ne \bar\Phi$ \cite{Dumitru:2005ng} generates
two separate but close values of inflection points leading to different  $T_\Phi$ and $T_{\bar\Phi}$. 
In that case one can take the average of $T_\Phi$ and $T_{\bar\Phi}$ as the deconfinement transition 
temperature. For $\mu=150$ MeV, we found  $T_\Phi=166$ MeV and 
$T_{\bar\Phi}=160$ MeV  and the average of the them ($163$ MeV)  is very close to 
the value of $T_\sigma=167$ MeV. With the  increase of $\mu$  the transition temperatures further get reduced; 
for example at $\mu=200$ MeV, $T_\Phi=153$ MeV, $T_{\bar\Phi}=151$ MeV and $T_\sigma= 153$ MeV.

\begin{figure}[hbt]
\includegraphics[scale=0.8]{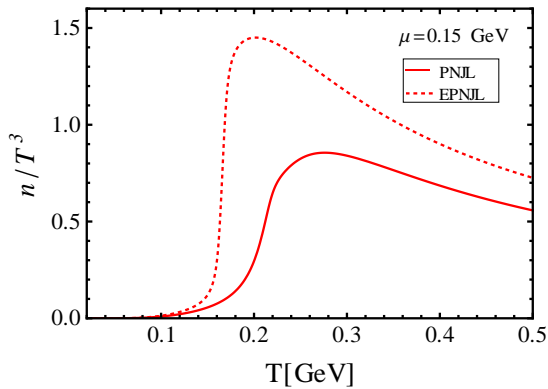}
\caption{Comparison of variations of scaled quark number density  between PNJL and EPNJL model for  $\mu=0.15$ GeV. }
\label{numberdensity_Gv0}
\end{figure}

We now discuss the differences in quark number density in EPNJL model with that of the PNJL one. 
In Fig.~\ref{numberdensity_Gv0} we observe that for temperature beyond 150 MeV the quark number density rises 
very sharply for EPNJL model as compared to PNJL one. This can be understood from Fig.~\ref{mass_fields_Gv0_Mu0} 
in which the value of Polyakov loop field rises very sharply beyond $T=150$ MeV for EPNJL model. 
This indicates that the Polyakov loop field provides a strong correlation among the quarks at low $T$ whereas 
the strength of the correlation among the quarks decreases when the value of the Polyakov loop field increases 
at high $T$ and we have more and more free quarks in the system for EPNJL model as compared to the PNJL one.

\subsubsection{With the isoscalar-vector interaction ($G_V\neq0$)}

Now we deal with the same set up but the isoscalar-vector interaction (${\tilde G}_V$) is turned on 
through (\ref{entangle_Gv}). In EPNJL model both couplings in (\ref{entangle_Gs}) and (\ref{entangle_Gv}) 
are entangled  and run with $T$ and $\mu$ by virtue of the gap equation in (\ref{gap_equation}).
We choose three different values of the strength of the isoscalar-vector interaction to demonstrate its 
effects within the EPNJL model. These values are taken in terms of $\tilde{G}_S$ and the reason for which is 
already mentioned in the sec.~\ref{Eff_QCD_model}.  


\begin{figure}[hbt]
\subfigure[]
{\includegraphics[scale=0.8]{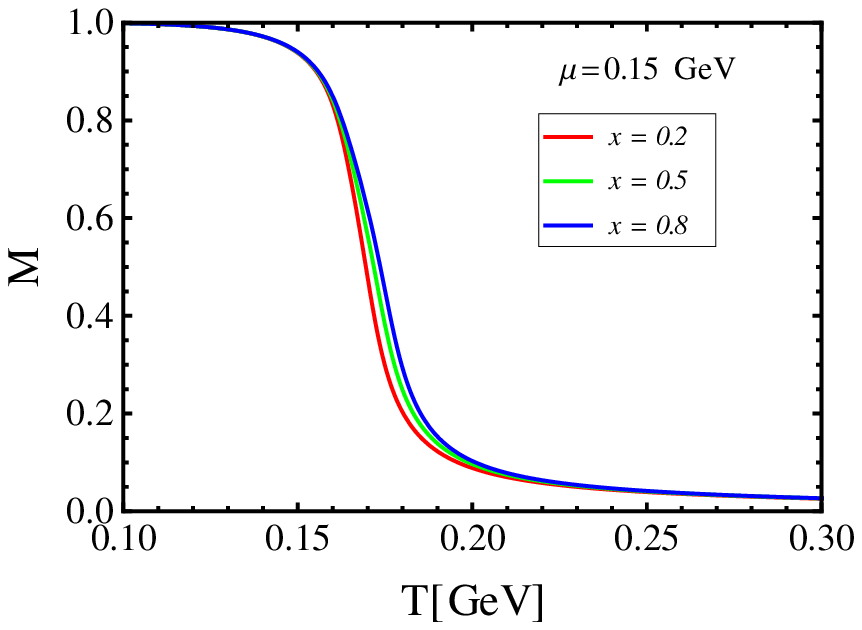}
\label{mass_differentGv}}
\subfigure[]
{\includegraphics[scale=0.785]{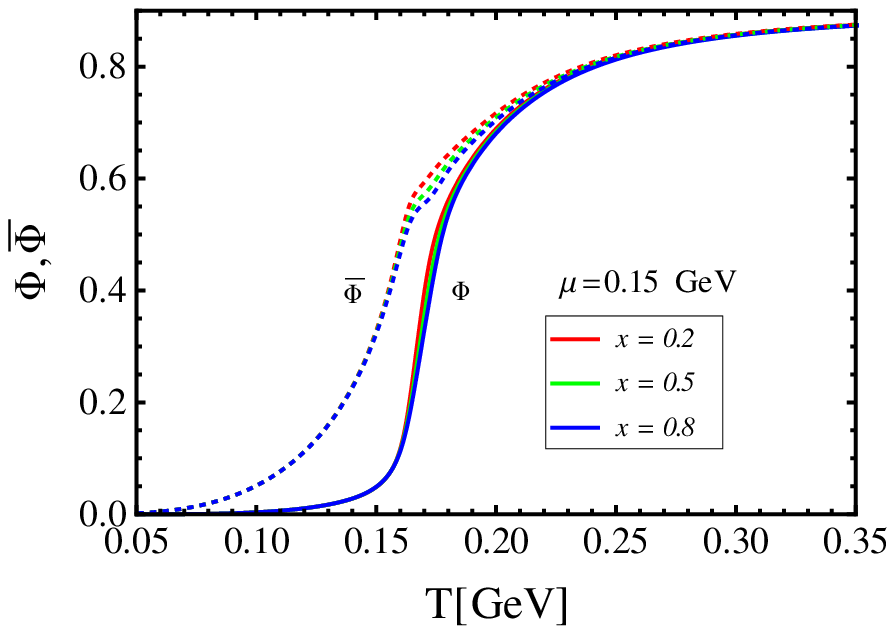}
\label{fields_differentGv}}
\caption{Variations of (a) scaled constituent quark mass and (b) Polyakov loop fields with temperature for 
three different values of $G_V$ at $\mu=0.15$ GeV in EPNJL model.}
\label{mass_fields_differentGv}
\end{figure}

In Fig.~\ref{mass_differentGv} the variation of the scaled constituent quark mass 
is shown for $\mu = 150$ MeV.  As one increases the strength of the vector interaction the rate of mass 
variation with the temperature  becomes slower.  
Since the couplings run in the EPNJL model the effect of the vector interaction is more prominent 
than that of the PNJL model with fixed values of couplings~\cite{Islam:2014sea}.
In the right panel 
(Fig.~\ref{fields_differentGv}) the variations of the Polyakov loop fields with temperature at $\mu = 150$ MeV are shown. 
We observe that with the increase of the value of $G_V$ the rate of increase of Polyakov loop fields with temperature 
decreases. The differences in the constituent quark masses or the Polyakov loop fields for different values of $G_V$ are however
more prominent within the temperature range $165\leq T(\mathrm {MeV}) \leq210$.

We have already discussed the effects of chemical potential on the transition temperatures in the previous section. Here in 
Table~\ref{table_Tsigma_Tphi} we present the variations of 
the transition temperatures by the inclusion of the vector interaction. It shows that as we increase 
the strength of the vector interaction for the same chemical potential, the values of $T_\sigma$ and as well as the average of $T_\Phi$ 
and $T_{\bar\Phi}$ increase \cite{Friesen:2014mha}.

\begin{table}
\begin{center}
\begin{tabular}{|cc|cccccc|}
 \hline
 & Values of $G_V$ and $\mu$ && $T_\Phi$ && $T_{\bar\Phi}$ & $\frac{T_\Phi+T_{\bar\Phi}}{2}$ & $T_\sigma$\\
 \hline
  & $x=0$,~~ $\mu=150$ MeV && 166 && 160  & 163  & 167\\
 \hline
  & $x=0.2$, $\mu=150$ MeV && 167 && 160  & 163.5  & 170\\
 \hline
  & $x=0.5$, $\mu=150$ MeV && 168 && 159  & 163.5  & 173\\
 \hline
  & $x=0.8$, $\mu=150$ MeV && 169 && 159  & 164  & 175\\
 \hline
 
\end{tabular}
\end{center}
\caption{Values of $T_\sigma$, $T_\Phi$ and $T_{\bar\Phi}$ for different values of $G_V$ and $\mu=150$ MeV.}
\label{table_Tsigma_Tphi}
\end{table}

\begin{figure}[hbt]
\subfigure[]
{\includegraphics[scale=0.8]{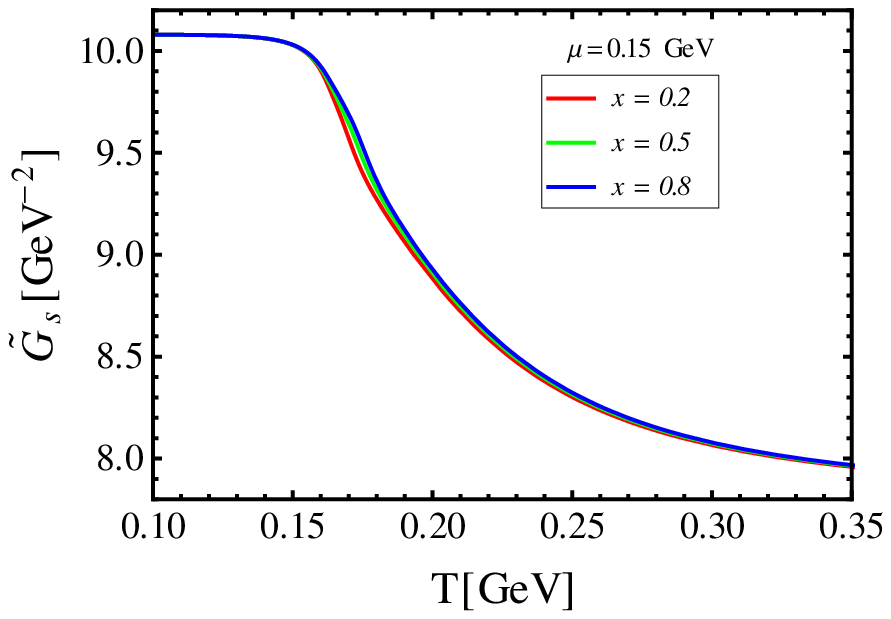}
\label{couplings_differentGv}}
\subfigure[]
{\includegraphics[scale=0.785]{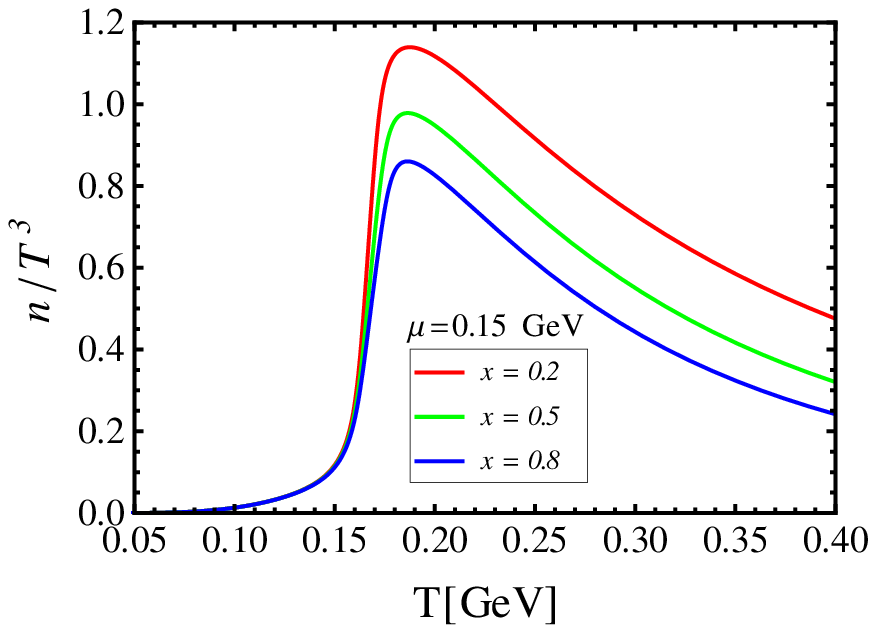}
\label{numberdensity_differentGv}}
\caption{Variations of  (a) scalar type four-quark coupling strength  and (b) scaled number density with 
temperature for three different values of $G_V$ at $\mu=0.15$ GeV in EPNJL model.}
\label{numberdensity_couplings_differentGv}
\end{figure}

In Fig.~\ref{couplings_differentGv} the variation of $\tilde{G}_S$ with temperature at ${\mu}=150$ MeV is shown for 
different value of $G_V$. It is found that the value of $\tilde{G}_S$ increases 
as the strength of the vector interaction increases for a given value of temperature and chemical potential. 
This can be understood from figure \ref{fields_differentGv} where the values of PL fields decrease
with increase of $G_V$. This in turn  leads to an enhancement of $\tilde{G}_S$ according to (\ref{entangle_Gs}).
In Fig.~\ref{numberdensity_differentGv} the variation of scaled quark number density with temperature is displayed for
same ${\mu}$ and  $G_V$ as in Fig.~\ref{couplings_differentGv}. For a given temperature and chemical potential the number density
is found to decrease with the increase of $G_V$. This is because the number of free quarks in the system is reduced
since the correlation among quarks increases due to the decrease  of PL fields  with  the increase of the couplings.

\subsection{Vector spectral function and dilepton rate}
\subsubsection{Without the isoscalar-vector interaction ($G_V=0$)}

\begin{figure}[hbt]
\subfigure[]
{\includegraphics[scale=0.8]{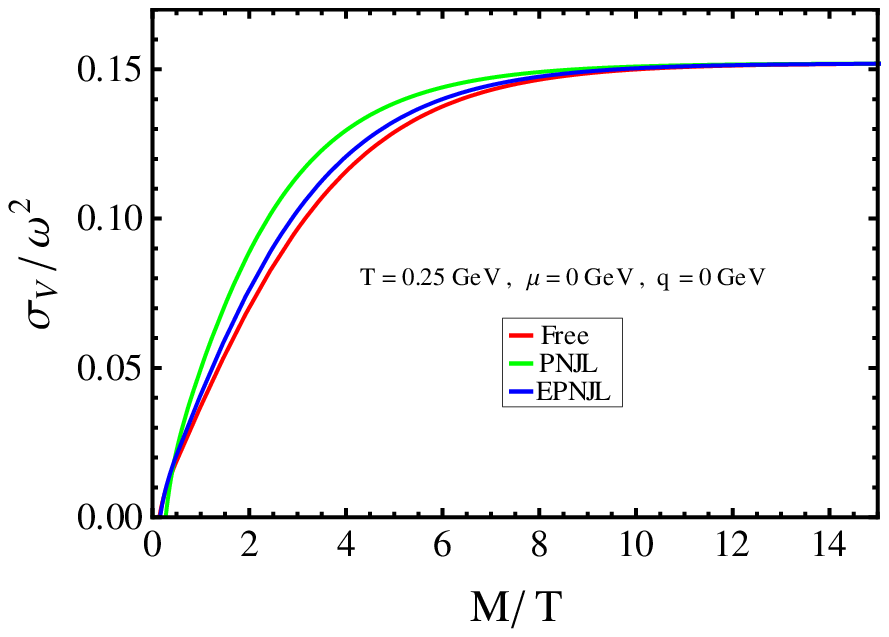}
\label{spectral_T0.25_Mu0}}
\subfigure[]
{\includegraphics[scale=0.8]{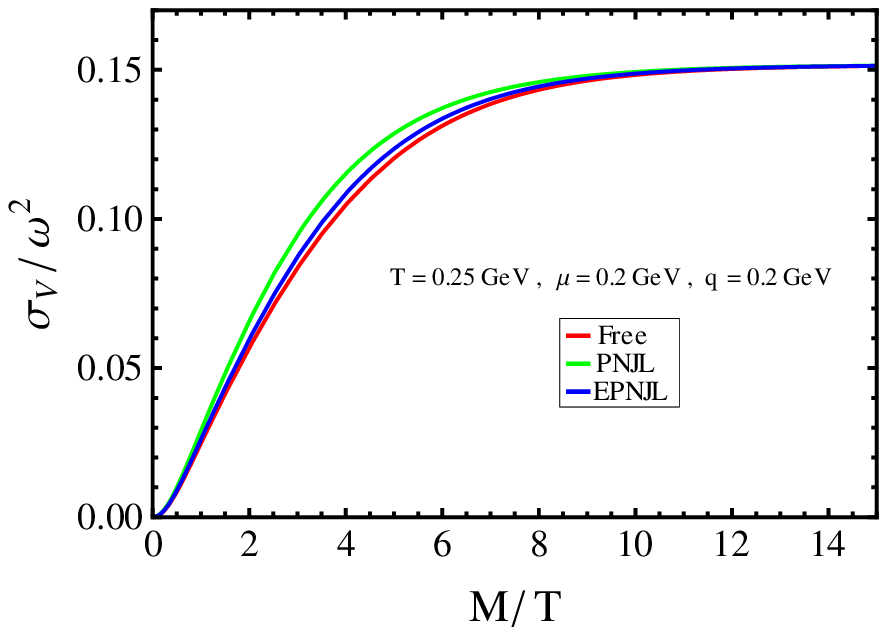}
\label{spectral_T0.25_Mu0.2}}
\caption{Scaled spectral function in PNJL and EPNJL model are compared with the free case for 
(a) $\mu=0$ and the three momentum $q=0$  and (b) $\mu=q=0.2$ GeV at $T=0.25$ GeV with $x=0$.}
\label{spectral_T0.25_Gv0}
\end{figure}

\begin{figure}[hbt]
\subfigure[]
{\includegraphics[scale=0.8]{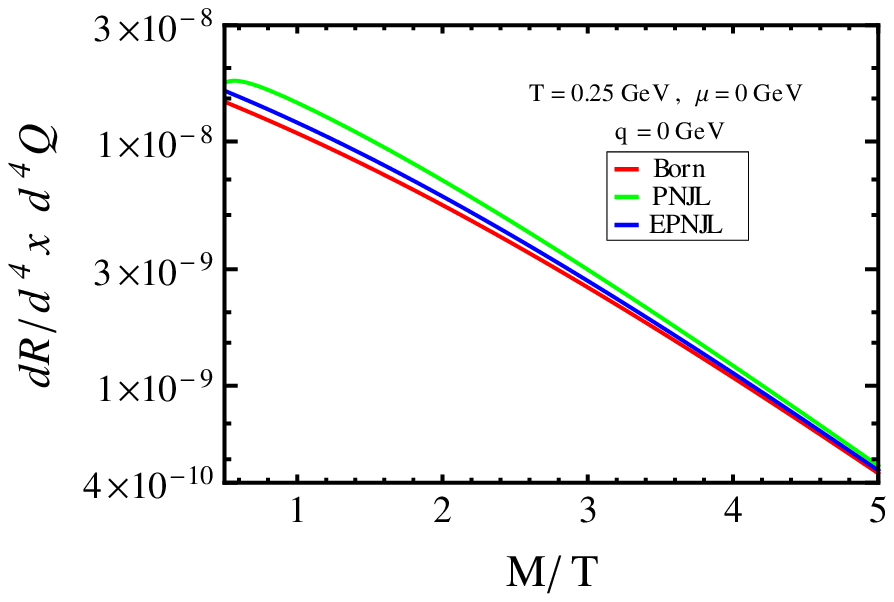}
\label{dilepton_T0.25_Mu0}}
\subfigure[]
{\includegraphics[scale=0.8]{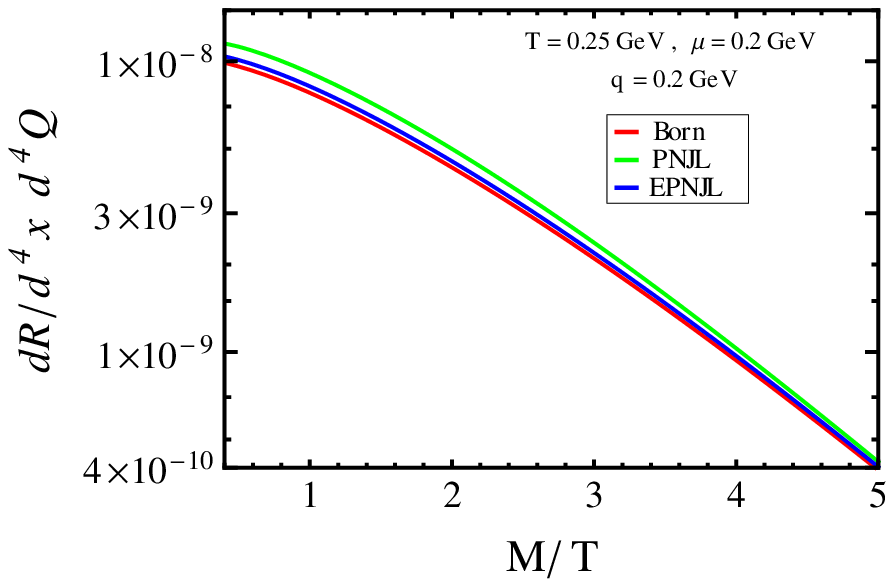}
\label{dilepton_T0.25_Mu0.2}}
\caption{Dilepton rate as a function of $M/T$ at $T=0.25$ GeV for PNJL and EPNJL model at (a) $\mu=q=0$ and (b) $\mu=q=0.2$ GeV with $x=0$.
The leading order perturbative dilepton (Born) rate is also shown.}
\label{dilepton_T0.25_Gv0}
\end{figure}

Now we will be discussing the entanglement effect on the spectral function vis-a-vis the dilepton rates without the inclusion 
of the vector interaction but considering only the scalar type interaction. In  Fig.~\ref{spectral_T0.25_Mu0} the spectral functions 
with zero external momentum ($q$) and zero chemical potential ($\mu$) i.e., $q=\mu=0$, for PNJL and EPNJL model along with the free 
case are displayed  whereas those in  Fig.~\ref{spectral_T0.25_Mu0.2} are for $q=\mu=200$ MeV. The corresponding dilepton rates are 
shown in Fig.~\ref{dilepton_T0.25_Gv0}. Due to the entanglement effect through scalar type interaction the spectral function vis-a-vis dilepton 
rate for EPNJL model gets suppressed  compared to PNJL model but is still higher than the Born rate. This could be understood
in the following way.  Usually the color degrees of freedom are suppressed in PNJL model due to the nonperturbative effect of 
the Polyakov Loop that causes an enhancement~\cite{Islam:2014sea} of the dilepton rate compared to the Born one. As soon as the 
entanglement effect is introduced through the scalar type interaction that relatively enhances the color degrees of freedom 
in the system due to the running in ${\tilde G}_S$ as evident from Fig.~\ref{numberdensity_Gv0}, hence the dilepton rate is reduced 
compared to that in PNJL model.

\subsubsection{With the isoscalar-vector interaction ($G_V\neq0$)}

\begin{figure}[hbt]
\subfigure[]
{\includegraphics[scale=0.8]{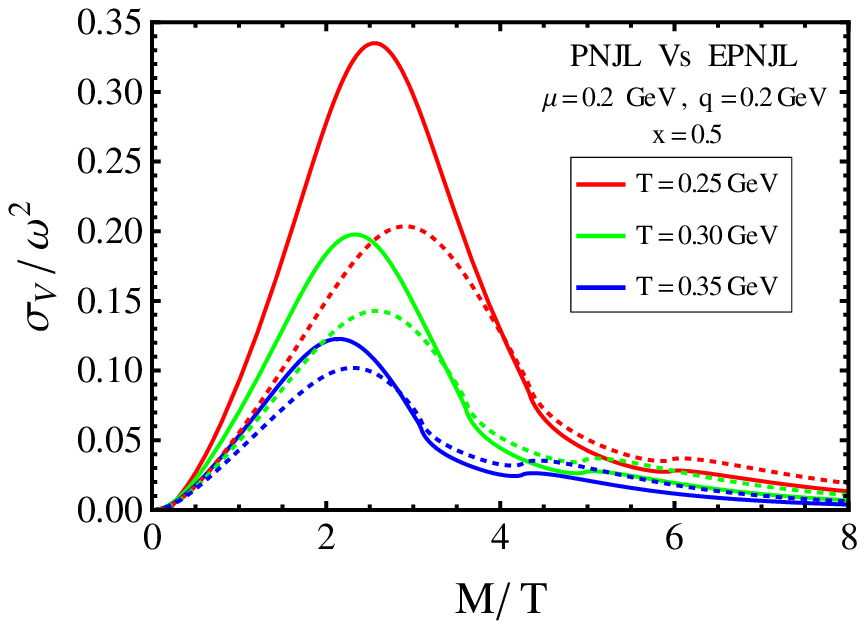}
\label{spectral_differentT}}
\subfigure[]
{\includegraphics[scale=0.78]{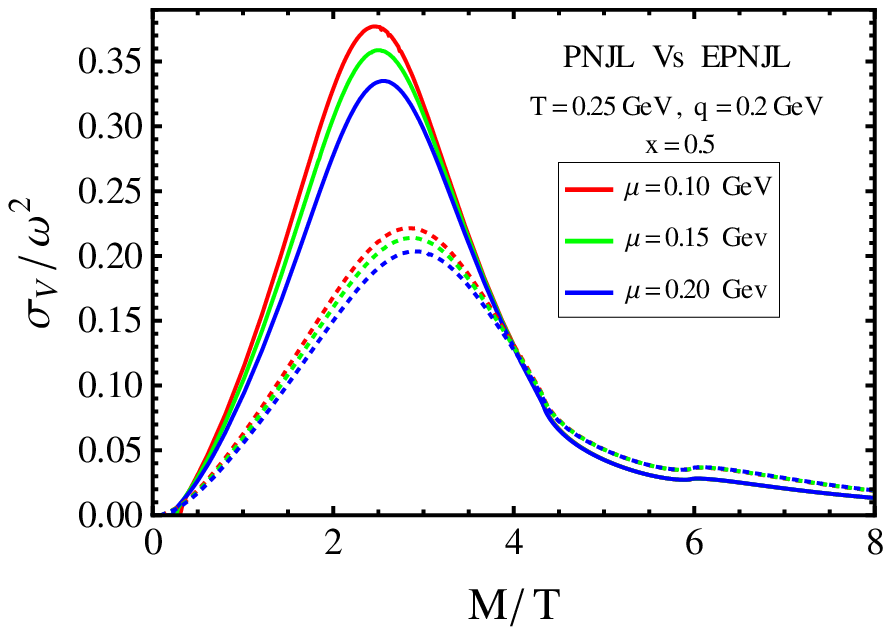}
\label{spectral_differentMu}}
\caption{Comparison between scaled spectral functions in PNJL (solid lines) and EPNJL (dotted lines) model for 
(a) a given chemical potential but different temperatures and (b) a given temperature but different chemical potentials 
at $x = 0.5$.}
\label{spectral_differentTMu}
\end{figure}

The free spectral function, in general, has a peak that appears at infinite value of $M$. This is also true for
four-quark scalar type interaction as seen above. However, in presence of isoscalar-vector 
interaction $G_V$ the peak appears at finite $M$ in the resummed spectral function in (\ref{eq.spectral_resum}) for given $G_V$ and $T$:
(a) below the kinematic threshold, $M <2M_f$, the resummed spectral function has a $\delta$-like peak due to the pole 
that can lead to bound state information of the vector meson and
(b) above the threshold  $M > 2M_f$ the resummed spectral function  picks
up a continuous contribution along with a somewhat broader peak\footnote{The width of the peak will depend on the value of $T$. If $T$ is around $T_c$ the peak will still be sharp around the threshold~\cite{Islam:2014sea}.}. We here concentrate on the continuous contribution ($M > 2M_f$) 
of the spectral function above $T_c$ that provides a finite width to a vector meson which decays to lepton pairs.
Now we focus on the effects of entanglement on spectral function and dilepton rate when the vector interaction is included
in addition to the scalar type interaction. In the left panel (Fig.\ref{spectral_differentT}) the scaled spectral functions at $\mu=200$ MeV
in PNJL (solid line) and EPNJL (dotted line) model  are shown for three different values of $T$. 
In Fig.\ref{spectral_differentT} the peak of the vector spectral function, for a given $T$ and $G_V$, 
is found to be suppressed and shifted to a higher $M$ in EPNJL model compared to PNJL one. This is purely due to the entangled vector 
interaction as the correlation among the quarks in the deconfined states becomes weaker in EPNJL model. 
In particular, the suppression is larger at lower value of $T$ and becomes smaller with the increase of
$T$.

The right panel (Fig.\ref{spectral_differentMu}) displays the same quantity for three different values of $\mu$ but at a given
$T=250$ MeV. Comparison with the left panel reveals that the variation of the suppression of the spectral function due to entanglement is strongly 
temperature dependent than the chemical potential.

\begin{figure}[hbt]
\subfigure[]
{\includegraphics[scale=0.8]{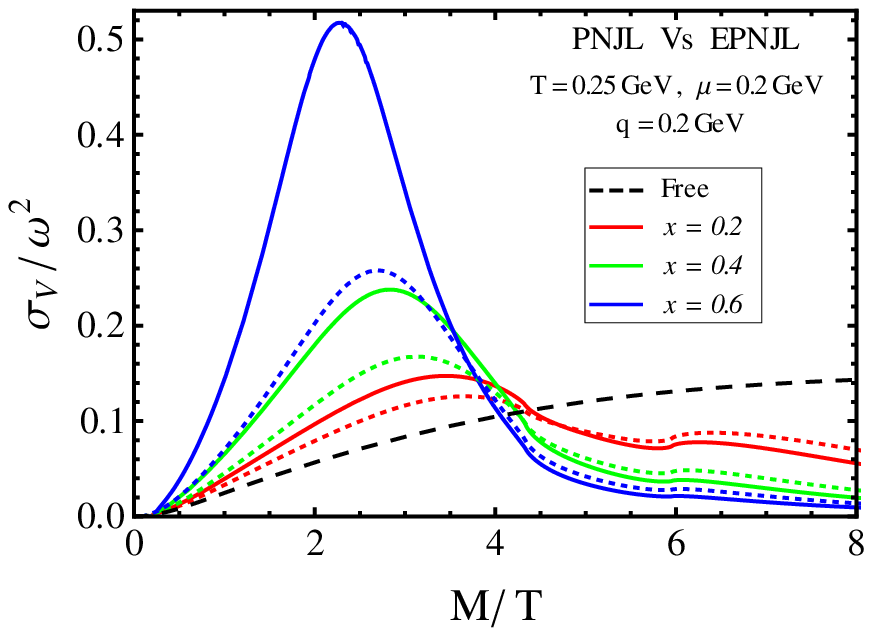}
\label{spectral_differentGv}}
\subfigure[]
{\includegraphics[scale=0.87]{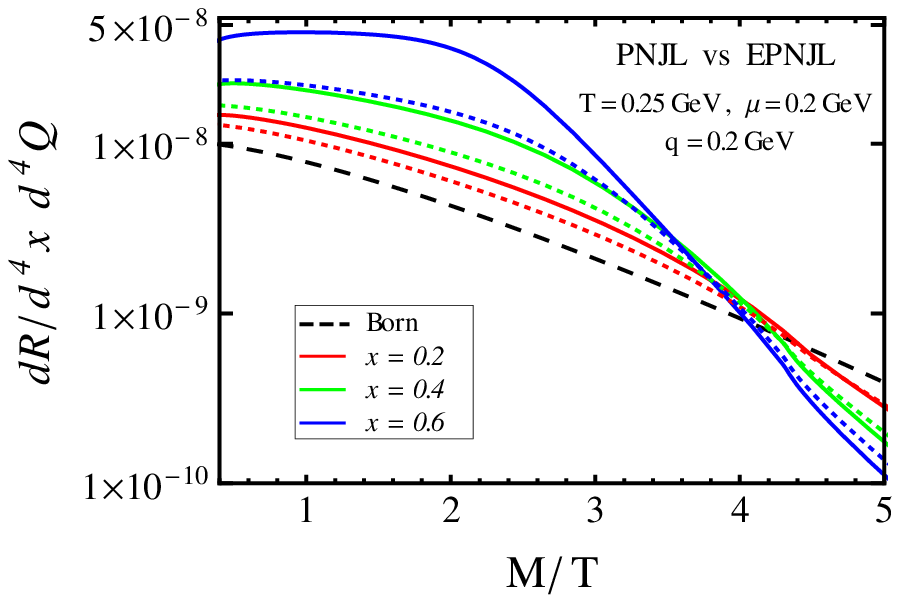}
\label{dilepton_differentGv}}
\caption{Plot of (a) scaled spectral function and (b) dilepton rate as a function of $M/T$ for PNJL (solid line) and EPNJL (dotted line) 
model for $T = 0.25$ GeV and $\mu=0.2$ GeV at three different choices of $G_V$.}
\label{spectral_dilepton_differentGv}
\end{figure}

In the left panel (Fig.~\ref{spectral_differentGv}) the spectral functions in PNJL (solid line) and EPNJL (dotted line) model 
at $T=250$ MeV and $\mu=200$ MeV for three different choices of $G_V$ are compared. As evident for any value of $G_V$ the strength 
of the spectral function for PNJL model is greater than that in the EPNJL one. In EPNJL model both couplings are strongly entangled 
through the mean fields and as one increases the strength of the vector interaction ($G_V$) that enhances the strength of the both 
running couplings. This in turn provides an enhancement in the strength of the spectral function that
decays to the dilepton pairs in the medium. The entanglement effect becomes more prominent than that with only the scalar 
type interaction. These features 
are well reflected in the right panel (Fig.~\ref{dilepton_differentGv}) where the corresponding dilepton rates in 
PNJL and EPNJL models are compared. For a given $G_V$ there is more lepton pairs at low mass in both EPNJL and PNJL model
compared to the leading order (Born) rate. Moreover, EPNJL model produces less lepton pairs than PNJL one. This is due to the entangled 
vector interaction that reduces the correlation among the quarks in the medium. However, as the strength of the vector interaction
increases, there is a relatively more dilepton production in both models.

\section{Conclusions}
\label{concl}

In general PNJL model contains nonperturbative information of confinement/deconfinement dynamics 
through the Polyakov Loop fields in addition to the chiral symmetry breaking dynamics. This model 
also employs the coupling of local scalar type four-quark interaction as well isoscalar-vector 
interaction. The scalar type four-quark coupling 
strength is fixed along with three momentum cutoff $\Lambda$ and bare quark mass $m_0$ to reproduce 
known zero temperature chiral physics in the hadronic sector. However, the value of the vector coupling 
is difficult to fix from the mass scale ($\rho$-meson mass) which is greater than the intrinsic 
scale $\Lambda$ of the effective theory. Some efforts have been made in the literature to estimate the 
value of the vector coupling mainly by fitting lattice data. However, there exists ambiguity about 
its value as discussed. Nevertheless, the introduction of vector interaction in heavy-ion physics 
is important for study of the spectral property like dilepton rate at non-zero chemical potential. 
On the other hand, in nuclear astrophysics the formation of stars with quark matter core depends
strongly on the existence of a quark vector repulsion. However, in PNJL model both the couplings are  
considered to be constant in the literature. Since, this model also contains temporal gluons 
these couplings, in principle, should depend on the Polyakov Loop fields. But this dependence is
usually neglected and the correlation of the Polyakov Loop and chiral dynamics  
is a weak one as it arises through the covariant derivative that couples the quark and 
the temporal gauge field in the model. 

In this article we have extended the usual PNJL model by introducing a strong entanglement between 
the chiral ($\sigma$) and the Polyakov Loop dynamics ($\Phi$), known as EPNJL model in the literature. 
The strong entanglement has been introduced  via effective four-quark scalar type  
interaction that obeys the centre symmetry, $Z(3)$ of pure $SU(3)$ gauge group. 
Since the Polyakov Loop and chiral fields run with temperature and chemical potential,
the entanglement  makes also those coupling run. This entanglement effect is capable of 
reproducing the coincidence of chiral ($T_\sigma$) and deconfinement ($T_\Phi$) transition 
temperature within the range provided by the 2-flavour lattice data.

The spectral function of the vector current-current correlation is related to 
the production of lepton pairs, which is considered as an important probe of the deconfined hadronic 
matter and has been measured in  high energy heavy-ion experiments~\cite{Adare:2009qk,Adler:2006yt}. 
On the other hand,  at RHIC and LHC energies the maximum temperature reached of a 
hot and dense strongly interacting matter created is not very far from the phase transition 
temperature $T_c$ and is nonperturbative in nature. In LQCD framework the dilepton production 
rate~\cite{Ding:2011} at finite temperature but zero chemical potential has also been computed  using a spectral 
function obtained from Euclidean correlation function through a probabilistic method that 
involves certain uncertainties and intricacies~\cite{Islam:2014sea}. In our previous calculation~\cite{Islam:2014sea}  
within PNJL model, the influence of the four-quark scalar and isoscalar-vector interaction without entanglement effect 
on the spectral function vis-a-vis the dilepton production was studied. In this article  
we have updated the spectral function and the dilepton production rate within the EPNJL 
model that takes into consideration the entanglement between the Polyakov loop and the chiral dynamics through 
scalar and vector interaction.  
In PNJL model both scalar and vector couplings do not run and the dominance of the Polyakov Loop 
fields substantially suppresses the color degrees of freedom around the phase transition temperature. On the other 
hand  EPNJL model introduces a strong entanglement between the chiral and the Polyakov loop dynamics which 
relatively enhances color degrees of freedom in the deconfined phase compared to the PNJL model. 
Because of this the strength of the vector spectral function is suppressed and the peak is shifted to a higher energy 
compared to that of  PNJL model but the strength is higher than the free one at low energy. 
Since  the dilepton production  is related to the vector spectral function, it is also suppressed in EPNJL model 
compared to the PNJL model but is more compared to the Born rate (leading order perturbative one) in the deconfined phase. 
This indicates relatively less production of lepton pairs at low energy with entangled vector interaction. However, 
as the strength of the vector interaction is increased there is a relative increase in the strength of the spectral 
function in both EPNJL and PNJL model, which also results in a relatively more production in lepton pairs at low invariant mass.

\begin{acknowledgments}
  Authors thankfully acknowledge the useful discussion with Rajarshi Ray, Anirban Lahiri, Peter Petreczky and Prasad Hegde
  during the course of this work.
  CAI would like to acknowledge the financial support from the University Grants Commission, India. SM and MGM 
  acknowledge the Department of Atomic Energy, India for the financial support through the project name $``$Theoretical 
  Physics Across the Energy Scale (TPAES)''.  
 \end{acknowledgments}

\end{document}